\documentclass[11pt]{article}

\usepackage{amsmath}
\usepackage{graphicx}
\usepackage{amsfonts}
\usepackage{amssymb}
\usepackage{epsfig}
\usepackage{color}
\usepackage{psfrag}
\usepackage{epstopdf}

\newcommand{\EQ}{\begin{equation}}
\newcommand{\EN}{\end{equation}}
\newcommand{\be}{\begin{equation}}
\newcommand{\ee}{\end{equation}}
\newcommand{\bea}{\begin{eqnarray}}
\newcommand{\eea}{\end{eqnarray}}

\newcommand{\nn}{\nonumber \\}
\DeclareMathOperator{\sgn}{sgn}
\DeclareMathOperator{\Arccos}{Arccos}

\begin{document} \setcounter{page}{0}
\topmargin 0pt
\oddsidemargin 5mm
\renewcommand{\thefootnote}{\arabic{footnote}}
\newpage
\setcounter{page}{0}
\topmargin 0pt
\oddsidemargin 5mm
\renewcommand{\thefootnote}{\arabic{footnote}}
\newpage
\begin{titlepage}
\begin{flushright}
\end{flushright}
\vspace{0.5cm}
\begin{center}
{\large {\bf On superuniversality in the $q$-state Potts model with quenched disorder}}\\
\vspace{1.8cm}
{\large Gesualdo Delfino$^{1,2}$ and Elena Tartaglia$^{1,2}$}\\
\vspace{0.5cm}
{\em $^1$SISSA -- Via Bonomea 265, 34136 Trieste, Italy}\\
{\em $^2$INFN sezione di Trieste}\\
\end{center}
\vspace{1.2cm}

\renewcommand{\thefootnote}{\arabic{footnote}}
\setcounter{footnote}{0}

\begin{abstract}
\noindent
We obtain the exact scale invariant scattering solutions for two-dimensional field theories with replicated permutational symmetry $\mathbb{S}_q$. After sending to zero the number of replicas they correspond to the renormalization group fixed points of the $q$-state Potts model with quenched disorder. We find that all solutions with non-zero disorder possess $q$-independent sectors, pointing to superuniversality (i.e. symmetry independence) of some critical exponents. The solution corresponding to the random bond ferromagnet, for which disorder vanishes as $q\to 2$, allows for superuniversality of the correlation length exponent $\nu$ \cite{random}. Of the two solutions which are strongly disordered for all values of $q$, one is completely $q$-independent and accounts for the zero-temperature percolation fixed point of the randomly bond diluted ferromagnet. The other is the main candidate to describe the Nishimori-like fixed point of the Potts model with $\pm J$ disorder, and leaves room for superuniversality of the magnetic exponent $\eta$, a possibility not yet excluded by available numerical data.
\end{abstract}
\end{titlepage}

\newpage

\section{Introduction}
The statistical systems with quenched disorder are those in which a part of the degrees of freedom (the disorder) does not reach thermal equilibrium on the timescale required for the other degrees of freedom, and can be treated stochastically. These systems display critical phenomena of the same qualitative type as the non-quenched ones, but are able to generate new critical exponents. The most common theoretical approach to critical properties has been provided by a treatment of disorder known as replica method which, for weak disorder, allows to cast the problem into the framework of the perturbative renormalization group (see e.g. \cite{Cardy_book}). This accounts for the Harris criterion \cite{Harris} for the appearance of disordered (or random) renormalization group fixed points, and in some cases yields perturbative results for the exponents. On the other hand, the determination of critical properties at strong disorder for systems with finite range interactions has essentially relied on numerical methods. In particular, the absence of exact solutions in the two-dimensional case, for which they are commonplace in pure systems, has been a peculiar, and in some respects mysterious, feature of the field. 

It has been known for some time that the two-dimensional $q$-state Potts model provides a particularly interesting playground for the study of critical behavior in presence of quenched disorder. Starting from the pure ferromagnet, weak bond randomness is marginally irrelevant at $q=2$ (Ising) \cite{DD} and can be studied perturbatively for $q\to 2$ \cite{Ludwig,DPP}. In addition, it is rigorously known \cite{AW} to turn the phase transition, which in the pure model is first order for $q>4$, into a second order one. Numerical simulations observed this phenomenon \cite{CFL} but also pointed to critical exponents preserving their Ising value for $q>2$ \cite{CFL,DW}. This {\it superuniversality}, which does not appear in the perturbative calculations, was also suggested for large $q$ by a simplified interfacial model \cite{KSSD}. The authors of this study observed that their analysis involves strong disorder and could refer to a line of fixed points different from the perturbatively accessible one. This possibility was also suggested in \cite{CJ}, where a numerical transfer matrix study found exponents depending on $q$ (weakly for the correlation length exponent $\nu$) and consistent with the perturbative results for $q\to 2$. Following numerical studies for large $q$ \cite{CB2,OY,JP}, exact values for the exponents at $q=\infty$ were conjectured in \cite{AdAI}, in particular the Ising value $\nu=1$. 
 
Recently it has been shown that the replica method can be implemented {\it exactly} in two dimensions at criticality \cite{random}, in the framework of scale invariant scattering theory \cite{paraf}. This approach yields exact solutions for the scattering amplitudes of the particles of the underlying conformal field theory. These solutions describe the renormalization group fixed points of the statistical system. For the Potts model, which is characterized by the permutational symmetry $\mathbb{S}_q$, it was found in \cite{random} that there exists and is unique a scattering solution which is defined for real values of $q\geq 2$ and corresponds to weak disorder for $q\to 2$. This solution interpolates from weak to strong disorder as $q-2$ grows from small to large values and, quite remarkably, is $q$-independent in the symmetry sector containing $\mathbb{S}_q$-invariant fields, thus allowing for $q$-independence of $\nu$ and $q$-dependence of the magnetic exponent $\eta$ \cite{random}. This scenario of partial superuniversality makes redundant the hypothesis of several critical lines for the random bond Potts ferromagnet, and will be further discussed in relation with the perturbative and numerical results in section~\ref{variable_disorder}.

In this paper we explore the full space of scale invariant scattering solutions for two-dimensional systems with quenched disorder and $\mathbb{S}_q$ symmetry. The analysis is expected to account for the possible fixed points of the Potts model with different disorder distributions. In this respect, it must be taken into account that, as recently shown for the pure model \cite{DT}, scale invariant scattering exhibits remarkable unifying properties: a single solution can account for fixed points with different exponents, so that a potentially infinite variety of cases (it is sufficient to think to the sensitivity of antiferromagnets to the lattice structure) admits a classification in terms of a relatively small number of scattering solutions. In the present case we find that all solutions with non-vanishing disorder possess $q$-independent sectors, in sharp contrast with the pure case where $q$-independence is absent \cite{DT}. In particular, we find only two solutions which are defined at $q=2$ and for which the disorder does not vanish as $q\to 2$. One of them is completely $q$-independent and should describe the zero-temperature percolation fixed point of the randomly bond diluted ferromagnet. The other is $q$-independent in a sector which, under a duality hypothesis, contains the spin field. This hypothesis makes room for superuniversality of the magnetic exponent $\eta=2\beta/\nu$ for the strong coupling fixed point in the model with randomly distributed ferromagnetic and antiferromagnetic bonds ($\pm J$ disorder); at $q=2$ this fixed point is known as Nishimori point \cite{Nishimori}. This possibility is compared with available numerical data in section~\ref{strong_disorder}.

The paper is organized as follows. In the next section we recall how the $q$-state Potts model with quenched disorder is described within the scale invariant scattering formalism for replicated $\mathbb{S}_q$ symmetry, and how the main equations are obtained from the requirements of crossing symmetry and unitarity. The solutions of these equations are then given in section~3. Section~4 is devoted to the discussion of the physical properties of the solutions, also with reference to the available numerical data. A summary of the main results and some additional remarks (in particular about interfacial properties at phase coexistence) are given in section~5.

\section{Random bond Potts model and scale invariant scattering}
The $q$-state Potts model \cite{Wu} is defined by the lattice Hamiltonian
\EQ
{\cal H}=-\sum_{\langle i,j\rangle}J_{ij}\delta_{s_i,s_j}\,,\hspace{1cm}s_i=1,2,\ldots,q\,,
\EN
where $s_i$ is a variable located at site $i$, the sum runs over nearest neighboring sites, and $J_{ij}$ are bond couplings. The model is characterized by the symmetry $\mathbb{S}_q$ associated to permutations of the $q$ values (``colors") that the site variable can take. In two dimensions the pure ferromagnet ($J_{ij}=J>0$) exhibits a phase transition which is of the second order up to $q=4$ and becomes of the first order for $q>4$ \cite{Baxter,Baxter_square_AF}. For pure antiferromagnets ($J_{ij}=J<0$) the situation is much more complex, since the presence and the order of the transition depend on the lattice structure and must be studied case by case (see e.g. \cite{SS} for a summary of exact results for the most common planar lattices, and \cite{Jacobsen_thresholds,JS} for numerical studies of more general lattices). In this respect, the prediction of the scale invariant scattering approach that a second order transition associated to the spontaneous breaking of the symmetry $\mathbb{S}_q$ cannot occur (in absence of randomness) for $q>(7+\sqrt{17})/2=5.5615..$ (see \cite{DT} and below) is particularly interesting for its generality. 

The random bond model corresponds to the case in which the couplings $J_{ij}$ (the disorder degrees of freedom) are random variables drawn from a probability distribution $P(J_{ij})$, and the average over disorder is taken on the free energy
\EQ
\overline{F}=\sum_{\{J_{ij}\}}P(J_{ij})F(J_{ij})\,,
\EN
in order to take into account that, in the quenched case, the thermalization timescale of the disorder degrees of freedom is extremely larger than that of the site variables $s_i$. Theoretically, the average over disorder is usually dealt with through the replica method \cite{EA}. Since $F=-\ln Z$, where $Z=\sum_{\{s_i\}}e^{-{\cal H}/T}$ is the partition function, the formal relation
\EQ
\overline{F}=-\overline{\ln Z}=-\lim_{n\to 0}\frac{\overline{Z^n}-1}{n}
\EN
maps the problem onto that of $n\to 0$ identical replicas coupled by the disorder average. 

It was shown in \cite{random} that the replica method can be implemented {\it exactly} for systems at (second order) criticality in two dimensions, within the scale invariant scattering approach introduced in \cite{paraf}.  The relevance of scattering theory\footnote{See \cite{fpu} for an overview on field, particles and critical phenomena in two dimensions.} originates from the fact that a two-dimensional Euclidean field theory is the continuation to imaginary time of a relativistic quantum field theory with one space and one time dimension. On the other hand, fixed points of the renormalization group in ($1+1$)-dimensional quantum field theories exhibit some specific features. In the first place the particles are right or left movers with energy and momentum related as $p=e>0$ and $p=-e<0$, respectively. In addition, infinite-dimensional conformal symmetry \cite{DfMS} forces the scattering to preserve infinitely many conserved quantities, and then to be completely elastic (the initial and final states are kinematically identical). It follows that the scattering amplitude of a right mover with a left mover depends only on the center of mass energy, which is the only relativistic invariant. However, since this invariant is dimensionful, scale invariance and unitarity imply that the amplitude is actually energy-independent, a circumstance which gives a particularly simple form \cite{paraf,fpu} to the unitarity and crossing equations \cite{ELOP} satisfied by the scattering amplitudes. 

\begin{figure}
\begin{center}
\includegraphics[width=10cm]{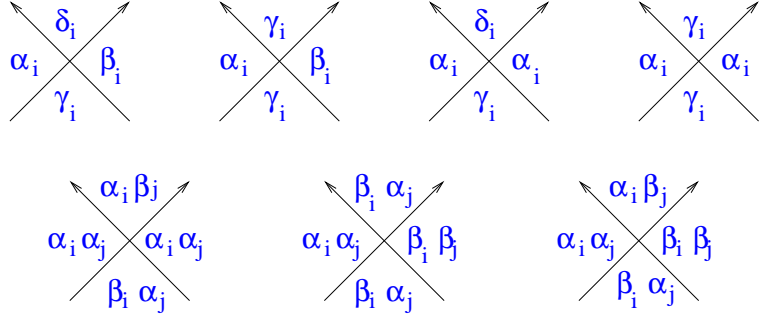}
\caption{Scattering processes corresponding to the amplitudes $S_0$, $S_1$, $S_2$, $S_3$, $S_4$, $S_5$, $S_6$, in that order. Time runs upwards. Different latin indices correspond to different replicas, and different greek letters for the same replica correspond to different colors. 
}
\label{potts_ampl}
\end{center} 
\end{figure}

The application of the scattering formalism to the Potts model relies on the fact that the latter is characterized by the permutational symmetry $\mathbb{S}_q$. A representation of this symmetry is carried by particles $A_{\alpha\beta}(p)$ ($\alpha,\beta=1,2,\ldots,q$; $\alpha\neq\beta$), which below the critical temperature describe the kinks which interpolate between pairs of degenerate ferromagnetic ground states \cite{CZ}. More generally, the space-time trajectories of the particles $A_{\alpha\beta}$ separate a region characterized by a color $\alpha$ from a region characterized by a color $\beta$, and these excitations provide the basic representation of permutational symmetry which applies also at criticality \cite{paraf}, where there are no kinks, and for antiferromagnets \cite{DT}. In order to extend the description to the random case, we have to consider the replicated theory with excitations $A_{\alpha_i\beta_i}$, where $i=1,2,\ldots,n$ labels the replicas \cite{random}. Now the trajectory of the particle $A_{\alpha_i\beta_i}$ separates a region characterized by colors $\alpha_1,\ldots,\alpha_n$ for replicas $1,\dots,n$, respectively, from a region where replica $i$ changed its color to $\beta_i$, with the colors of the other replicas unchanged. Then the requirement of invariance under permutations of the replicas and permutations of the colors within each replica ($\mathbb{S}_q\times\mathbb{S}_n$ symmetry) leaves us with the seven inequivalent amplitudes $S_0,S_1,\ldots,S_6$ depicted in figure~\ref{potts_ampl}, where we only keep track of the replicas whose color changes in the scattering process. The four amplitudes in the upper part of the figure have all particles in the same replica and are the amplitudes entering the pure case; the three remaining amplitudes introduce interaction among the replicas.

\begin{figure}
\begin{center}
\includegraphics[width=15cm]{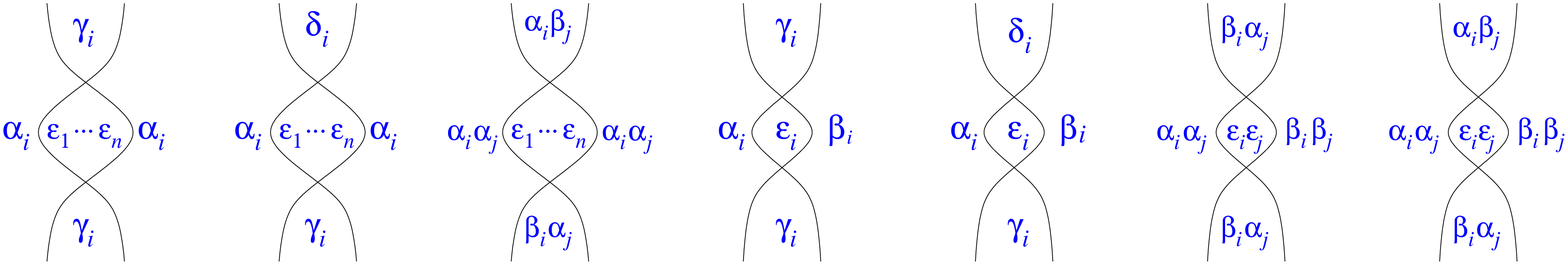}
\caption{Pictorial representations associated to the unitarity equations (\ref{u1})--(\ref{u6}), in the same order. The amplitude for the lower crossing multiplies the complex conjugate of the amplitude for the upper crossing, and sum over colors in the internal (i.e. closed) region is implied. Replicas which are not indicated keep the same color in the four external regions, and also in the internal region for the last four diagrams.
}
\label{random_unitarity}
\end{center} 
\end{figure}

In relativistic scattering theory \cite{ELOP} crossing symmetry establishes a relation between amplitudes which are exchanged under exchange of time and space directions. In the present case it takes the form
\bea
S_0 =S_0^* &\equiv & \rho_0\,,
\label{c1}\\
S_1 =S_2^*&\equiv &\rho\,e^{i\varphi}\,,
\label{c2}\\
S_3 =S_3^*&\equiv &\rho_3\,,
\label{c3}\\
S_4 =S_5^*&\equiv &\rho_4\,e^{i\theta}\,,
\label{c4}\\
S_6 =S_6^*&\equiv &\rho_6\,,
\label{c5}
\eea
where we introduced 
\EQ
\rho_0,\rho_3,\rho_6,\varphi,\theta\in\mathbb{R}\,,\hspace{1cm}\rho,\rho_4\geq 0\,.
\label{constraints}
\EN
Unitarity is a general property of the scattering matrix expressing conservation of probability. It is substantially simplified by complete elasticity and in the present case leads to the equations~\cite{random} 
\bea
&& \rho_3^2+(q-2)\rho^2+(n-1)(q-1)\rho_4^2=1\,,
\label{u1}\\
&& 2\rho\rho_3\cos\varphi+(q-3)\rho^2+(n-1)(q-1)\rho_4^2=0\,,
\label{u2}\\
&& 2\rho_3\rho_4\cos\theta+2(q-2)\rho\rho_4\cos(\varphi+\theta)+(n-2)(q-1)\rho_4^2=0\,
\label{u3}\\
&& \rho^2+(q-3)\rho_0^2=1\,,
\label{u4a}\\
&& 2\rho_0\rho\cos\varphi+(q-4)\rho_0^2=0\,,
\label{u4b}\\
&& \rho_4^2+\rho_6^2=1\,,
\label{u5}\\
&& \rho_4\rho_6\cos\theta=0\,,
\label{u6}
\eea
which correspond to the diagrams of figure~\ref{random_unitarity}. Equation (\ref{u5}) implies
\EQ
\rho_4=\sqrt{1-\rho_6^2}\in[0,1]\,.
\label{rho46}
\EN
It is relevant to note that $q$ and $n$ appear in the equations as parameters which can take real values. The possibility of analytic continuation in $q$ is known already from the lattice and allows to obtain percolation \cite{SA} in the limit $q\to 1$ \cite{FK} of the pure ferromagnet; a series of results obtained through this analytic continuation in the scattering framework, in particular for percolation, can be found in \cite{DV_3point,DVC,DV_4point,DV_crossing,DG_confinement}. On the other hand, analytic continuation in $n$ is essential for the replica approach. It is worth stressing that the unitarity codified by equations (\ref{u1})--(\ref{u6}) is that of the scattering matrix, and expresses the general fact that the probabilities of all scattering events with a given initial state sum to 1. The unitarity of the scattering matrix has no relation with reflection positivity (often called ``unitarity'' in conformal field theory); reflection positivity refers to positivity of two-point correlators and is lost for $q\neq 2,3,\ldots$ and/or $n\neq 1,2,\ldots$. Systems with quenched disorder correspond to $n\to 0$ and are not reflection positive.

As required, the unitarity equations (\ref{u1})--(\ref{u6}) reduce to those of the pure model \cite{paraf,DT} when $n$ is set to 1 and the equations which still involve $\rho_4$ and/or $\rho_6$ are ignored.  Another important property is that $\rho_4=0$ gives $n$ non-interacting replicas, since it corresponds to $S_4=S_5=0$ and, as a consequence of (\ref{u5}), to $S_6=\pm 1$. We recall that in one spatial dimension scattering leads two colliding particles to exchange their positions on the line, in such a way that a scattering amplitude equal to 1 (resp. $-1$) corresponds to non-interacting bosons (resp. fermions). 

The amplitudes of figure~\ref{potts_ampl} show how in general different final states can originate from the same initial state. It is possible, however, to build superpositions of two-particle states which diagonalize the scattering, i.e. scatter into themselves with amplitudes which, by unitarity, must be phases. In particular, the superposition  $\sum_{i,\gamma_i}A_{\alpha_i\gamma_i}A_{\gamma_i\alpha_i}$ scatters into itself with the phase
\EQ
S=S_3+(q-2)S_2+(n-1)(q-1)S_4\,.
\label{S}
\EN
We call ``neutral" this scattering channel involving excitations beginning and ending with the same color. In a similar way, the ``charged" superpositions $\sum_{\gamma_i}A_{\alpha_i\gamma_i}A_{\gamma_i\beta_i}$ and $A_{\alpha_i\beta_i}A_{\alpha_j\beta_j}+A_{\alpha_j\beta_j}A_{\alpha_i\beta_i}$ diagonalize the scattering with phases
\bea
\Sigma &=& S_1+(q-3)S_0\,,
\label{Sigma}\\
\tilde{\Sigma} &=& S_5+S_6\,,
\label{sigmatilde}
\eea
respectively.

It is known in general \cite{paraf} that a neutral phase is related to the conformal dimension $\Delta_\eta$ of the field $\eta$ which creates the massless particles as
\EQ
S=e^{-2i\pi\Delta_\eta}\,.
\label{phase}
\EN
We recall that a generic scaling field $\Phi$ has conformal dimensions $(\Delta_\Phi,\bar{\Delta}_\Phi)$ which determine its scaling dimension and spin as $X_\Phi=\Delta_\Phi+\bar{\Delta}_\Phi$ and $s_\Phi=\Delta_\Phi-\bar{\Delta}_\Phi$ respectively; the field $\eta$ is chiral ($\bar{\Delta}_\eta=0)$. Bosonic and fermionic particles are obtained when $\Delta_\eta$ is integer and half-integer, respectively, and particles with generalized statistics otherwise\footnote{In the off-critical case, generalized statistics does not affect low energy phenomena (see e.g. \cite{wedge}), but shows up in high energy asymptotics \cite{Smirnov,sqf,DN_ttbar}.}.

\section{Solutions of the unitarity equations}
The solutions of the unitarity equations (\ref{u1})--({\ref{u6}) satisfying (\ref{constraints}) and (\ref{rho46}) correspond to renormalization group fixed points possessing the symmetry $\mathbb{S}_q\times\mathbb{S}_n$. It is a general consequence of the form of the crossing and unitarity equations that, given a solution, another solution is obtained reversing the sign of all the amplitudes; correspondingly, all the solutions we write below admit a choice between upper and lower signs.  

\subsection{The pure case ($n=1$)}
For future reference we list in table~\ref{solutions} the solutions for the pure case \cite{DT}. As we said, this is obtained setting $n=1$ and ignoring the equations which still contain $\rho_4$ and/or $\rho_6$. It was shown in \cite{DT}  how solution III$_-$ is able to describe at the same time the critical and tricritical lines of the Potts ferromagnet, and the critical line of the square lattice Potts antiferromagnet. Solution I admits a description as a line of fixed points (parametrized by $\varphi$) with central charge $c=1$ (see \cite{DT}). The other solutions leave room for future identifications. In particular, solution V shows that a $\mathbb{S}_q$-invariant fixed point can be found up to the maximal value $q=(7+\sqrt{17})/2=5.5615..$, which is larger than the usually assumed maximal value $4$. Since the matter of $\mathbb{S}_q$ spontaneous breaking in ferromagnets is settled by universality, a realization of solution V, in particular for $q=5$, should be looked for in antiferromagnets, with infinitely many lattice choices available and inequivalent; candidates suggested by existing numerical data include the lattices studied in \cite{Huang,Deng}. 

\begin{table}
\begin{center}
\begin{tabular}{|c|c||c|c|c|c|}
\hline
Solution & Range & $\rho_0$ & $\rho$ & $2\cos\varphi$ & $\rho_3$ \\
\hline
I & $q=3$ & $0$, $2\cos\varphi$ & $1$ & $\in[-2,2]$ & $0$ \\ 
& & & & & \\
II$_\pm$ & $q\in[-1,3]$ & $0$ & $1$ & $\pm\sqrt{3-q}$ & $\pm\sqrt{3-q}$\\
& & & & & \\
III$_\pm$ & $q\in[0,4]$ & $\pm 1$ & $\sqrt{4-q}$ & $\pm\sqrt{4-q}$ & $\pm (3-q)$\\
& & & & & \\
IV$_\pm$ & $q\in[\frac{1}{2}(7-\sqrt{17}),3]$ & $\pm\sqrt{\frac{q-3}{q^2-5q+5}}$ & $\sqrt{\frac{q-4}{q^2-5q+5}}$ & $\pm\sqrt{(3-q)(4-q)}$ & $\pm \sqrt{\frac{q-3}{q^2-5q+5}}$\\
& & & & & \\
V$_\pm$ & $q\in[4,\frac{1}{2}(7+\sqrt{17})]$ & $\pm\sqrt{\frac{q-3}{q^2-5q+5}}$ & $\sqrt{\frac{q-4}{q^2-5q+5}}$ & $\mp\sqrt{(3-q)(4-q)}$ & $\pm \sqrt{\frac{q-3}{q^2-5q+5}}$ \\
\hline
\end{tabular}
\caption{Solutions of Eqs.~(\ref{u1}), (\ref{u2}), (\ref{u4a}) and (\ref{u4b}) with $n=1$. They correspond to renormalization group fixed points of $S_q$-invariant theories (without disorder).} 
\label{solutions}
\end{center}
\end{table}

\subsection{$n$ generic}
\label{n}
Coming to solutions with $n$ generic, (\ref{u6}) shows that we can distinguish two classes of solutions with coupled replicas ($\rho_4\neq 0$), that with $\cos\theta=0$ and that with $\rho_6=0$, which we now consider separately.

\subsubsection{$\cos\theta=0$}
\label{ct0}
Up to sign doubling, there are four solutions defined in intervals of $n$ which include $n=0$. For all of them the maximal value of $n$ is $\tfrac{2}{q-1}$ for\footnote{The fact that solutions allowing the analytic continuation to $n=0$ are defined for $n<2$ when $q>2$ is of interest for the perturbative studies of coupled Potts models (see \cite{DJLP} and references therein).} $q>2$. 

\noindent
The first solution is defined for $q\geq 3$ and reads
\begin{subequations}
\begin{align}
\rho_0 &=\rho_3=0, & \rho &=1, &\rho_4 &=\sqrt{\frac{3-q}{(n-1)(q-1)}}, \nonumber\\
&&&& 2\cos\varphi  &= \pm \frac{1}{q-2}\sqrt{\frac{n^2(q-1)(q-3)}{n-1}+4}.
\label{eq:quenched_ue_sol1}
\end{align}

\noindent
The second solution is defined for $q>\sqrt{2}$ and reads
\begin{align}
\rho_0&=0, & \rho&=1,&
\rho_3 &=2\cos\varphi = \pm\sqrt{\frac{n^2(q-1)(q-3)+4(n-1)}{(n-q)(n-nq+q)}},\nn
& & & &\rho_4 &=|q-2|\sqrt{\frac{q+1}{(q-1)(q-n)(n-nq+q)}}.
\label{eq:quenched_ue_sol2}
\end{align}

\noindent
The third solution is defined for $q\in[\tfrac{1}{2} (7-\sqrt{17}),4]$ and $q\geq\tfrac{1}{2} (7+\sqrt{17})$, and reads
\begin{align}
\begin{split}
\rho _0&=\pm\sqrt{\frac{(q-3)(q-1) n^2+4(n-1)}
{(q-1) (q^2-5 q+5) n^2+\left(q^2-6 q+6\right)^2 (n-1)}},\\
\rho &=\sqrt{\frac{(q-4)[(q-1)n^2+\left(q^3-8 q^2+16 q-12\right)(n-1)]}{(q-1) (q^2-5 q+5) n^2+\left(q^2-6 q+6\right)^2 (n-1)}},\\
\rho _3&=\pm\sqrt{\frac{(q-3)(q-1) n^2+4(n-1)}
{(q-1) (q^2-5 q+5) n^2+\left(q^2-6 q+6\right)^2 (n-1)}},\\
\rho _4&=\frac{|q-2|}{q-1} \sqrt{-\frac{(q-4)(q-1) (q^2-7 q+8)}{ (q^2-5 q+5) n^2+\left(q^2-6 q+6\right)^2 (n-1)}}, \\
 2 \cos\varphi &=\mp(q-4)\sqrt{\frac{(q-3)(q-1) n^2+4(n-1)}{(q-4)[(q-1)n^2+\left(q^3-8 q^2+16 q-12\right)(n-1)]}} .
\label{eq:quenched_ue_sol3} 
\end{split}
\end{align} 

\noindent
For positive values of $q$, the fourth solution is defined in the ranges $q\in[\tilde{q}= 1.488..,3]$ and $q\geq 4$, and reads
\begin{align}
\begin{split}
\rho _0&=\pm\sqrt{\frac{(q-3)(q-1) n^2+4(n-1)}{(q-3)(q-1) n^2+(q^2-4 q+2)^2(n-1)}},\\
\rho &=\sqrt{\frac{(q-4) \left[-(q-3)(q-1)n^2+
\left(q^3-4 q^2+4 q-4\right)(n-1)\right]}{(q-3)(q-1) n^2+\left(q^2-4 q+2\right)^2 (n-1)}},\\
\rho _3&=\mp(q-3) \sqrt{\frac{(q-3)(q-1)n^2+4(n-1)}{(q-3)(q-1)
n^2+\left(q^2-4 q+2\right)^2 (n-1)}},\\
\rho _4&=\frac{|q-2|}{q-1}\sqrt{-\frac{(q-4)(q-3)(q-1)q}{(q-3)(q-1)n^2+\left(q^2-4q+2\right)^2 (n-1)}},\\
2\cos\varphi&=\mp(q-4)\sqrt{\frac{ (q-3)(q-1) n^2+4 (n-1)}{(q-4)[-(q-3)(q-1)n^2+
\left(q^3-4 q^2+4 q-4\right)(n-1)]}}.
\end{split}
\label{eq:quenched_ue_sol4} 
\end{align}
\end{subequations}
The value $\tilde{q}$ is a root of the equation $q^4-8q^3+20 q^2-14 q-2 = 0$; below $\tilde{q}$ the solution would not include the value $n=0$. 

For all solutions, it is understood that $\rho_6$ follows from $\rho_4$ through (\ref{rho46}). Defining 
\EQ
\textrm{sgn}(\sin\theta\sin\varphi)=(-1)^j\,,
\label{j}
\EN 
all four solutions correspond to $j=0$ below $q=2$, and to $j=1$ above.

\subsubsection{$\rho_6=0$} 
\label{rho60}
There are two solutions in this class which admit continuation to $n=0$. Both are defined for any $q$, have
\EQ
\rho _0=0,\hspace{1cm}\rho=\rho_4=1,\hspace{1cm}\rho_3=2\cos \varphi=\pm\sqrt{n (1-q)+2}\,,
\label{r60n}
\EN
and differ in the value of $\cos\theta$. The first solution has
\begin{subequations}
\EQ
\cos\theta=\cos\varphi\,,
\EN
and $j=1$ in (\ref{j}), while the second 
\EQ
2\cos\theta=\pm\frac{\sqrt{n (1-q)+2} \left(n (q-1)-q^2+2\right)}{\left(n (q-1)^2-q^2+2 q-2\right)}\,,
\EN
\end{subequations}
$j=0$ for $q\in[2-\sqrt{2},2+\sqrt{2}]$, and $j=1$ outside this range.

\subsection{The limit $n\to 0$}
\label{n0}
Specializing to $n=0$ the four solutions (\ref{eq:quenched_ue_sol1})-(\ref{eq:quenched_ue_sol4}) with $\cos\theta=0$ we obtain that the first of them, defined for $q\geq 3$, reads
\begin{subequations}
\begin{align}
\rho _0 &=\rho_3=0, & \rho &=1, & \rho _4&=\sqrt{\frac{q-3}{q-1}}, & 2 \cos\varphi &=\pm\frac{2}{q-2}\,;
\label{n0a}
\end{align}
the second, defined for $q>\sqrt{2}$, reads
\begin{align}
\rho _0&=0, & \rho &=1, & \rho _3&=2 \cos\varphi =\pm\frac{2}{q}, 
& \rho _4&=\frac{|q-2|}{q} \sqrt{\frac{q+1}{q-1}}\,;
\label{n0b}
\end{align}
the third, defined for $q\in[\tfrac{1}{2} (7-\sqrt{17}),4]$ and $q\geq\tfrac{1}{2} (7+\sqrt{17})$, reads
\begin{align}
\rho_0&=\pm\frac{2}{| q^2-6 q+6|}\,,  &
\rho &=\frac{\sqrt{(q-4) (q^3-8 q^2+16 q-12)}}{|q^2-6 q+6|}\,, \nn
\rho _3&=\pm\frac{2}{| q^2-6 q+6|}\,, &
\rho _4&=\frac{|q-2|\sqrt{(q-4)(q-1)(q^2-7 q+8)}}{(q-1)|q^2-6 q+6|}\,, \nn
2 \cos\varphi &=\mp\frac{2 (q-4)}{\sqrt{(q-4) (q^3-8 q^2+16 q-12)}}\,; &&
\label{n0c}
\end{align}
finally the fourth solution, which for $q$ positive is defined in the ranges $q\in[\tilde{q}= 1.488..,3]$ and $q\geq 4$, reads
\begin{align}
\rho _0&=\pm\frac{2}{|q^2-4 q+2|}\,, &
\rho &=\frac{\sqrt{(q-4) (q^3-4 q^2+4 q-4)}}{|q^2-4 q+2|}\,,\nn
\rho _3&=\mp\frac{2(q-3)}{| q^2-4 q+2| }\,, &
\rho _4&=\frac{|q-2|\sqrt{(q-4) (q-3) (q-1) q}}{(q-1)|q^2-4 q+2|}\,, \nn
2 \cos\varphi &=\mp\frac{2 (q-4)}{\sqrt{(q-4) (q^3-4 q^2+4 q-4)}}\,, &&
\label{n0d}
\end{align}
with $\rho_6=0$ at $\tilde{q}$.
\end{subequations}

For $n=0$ the two solutions with $\rho_6=0$, which are defined for any $q$, read
\begin{subequations}
\begin{align}
\rho _0&=0,&\rho&=\rho_4=1, &  \rho_3&=2\cos\varphi=2\cos\theta =\pm\sqrt{2}\,, 
\label{r60a}
\end{align}
and 
\begin{align}
\rho _0&=0,&\rho&=\rho_4=1, & \rho_3&=2\cos\varphi =\pm\sqrt{2}, & 
2\cos\theta &= \pm\frac{\sqrt{2}(q^2-2)}{\left(q^2-2 q+2\right)}\,, 
\label{r60b}
\end{align}
\end{subequations}

\section{Properties of the solutions}
We have noted above how the scattering solutions with $\rho_4=0$ correspond to decoupled replicas. Since the replicas are coupled by the average over disorder, the parameter $\rho_4\in[0,1]$ gives a measure of the disorder strength within the scattering formalism. It follows that the two classes of solutions obtained in the previous section, that with $\cos\theta=0$ and that with $\rho_6=0$, are physically distinguished by the dependence of the disorder strength on the parameter $q$. While the four solutions in the first class all admit weak disorder limits, the two solutions in the second class are strongly disordered for any $q$. We now separately discuss these two classes of solutions.

\subsection{Solutions with weak disorder limit}
\label{variable_disorder}
Focusing on the solutions with $\cos\theta=0$, it is interesting to consider the zeros of $\rho_4$, i.e. the values of $q$ for which $\rho_4$ vanishes. When $\rho_4=0$ the $n$-dependence drops out of the unitarity equations (\ref{u1})--(\ref{u6}), as it should for decoupled replicas. It follows that the zeros of $\rho_4$ must be $n$-independent and must fall in the union of the ranges of $q$ spanned by the solutions of the pure model (see table~\ref{solutions}), i.e. in the interval $[-1,\frac{7+\sqrt{17}}{2}]$; the solutions of section~\ref{ct0} explicitly show that this is the case\footnote{Notice that the polynomial $q^2-7 q+8$ appearing in solution (\ref{eq:quenched_ue_sol3}) coincides with $(q-\frac{7-\sqrt{17}}{2})(q-\frac{7+\sqrt{17}}{2})$.}.

It is interesting that the possible zeros of $\rho_4$ can be predicted from the knowledge of the solutions of the pure model and renormalization group considerations. To see this, observe that in a neighborhood of a zero, let us call it $q_0$, the replicas are weakly coupled, so that the solution should describe a perturbative random fixed point, namely a random fixed point which is sufficiently close to the pure fixed point to be perturbatively approachable. In the framework of the perturbative renormalization group (see e.g. \cite{Cardy_book}) this notion of closeness translates into the fact that the field which drives the flow between the pure and the random fixed point is almost marginal, and is marginal when the two fixed points coalesce, namely at $q_0$. The Harris criterion amounts to say that the scaling dimension of the field which drives the flow is twice the dimension $X_\varepsilon$ of the energy density field of the {\it pure} model, and in two dimensions a field is marginal if it has scaling dimension 2. It follows that $X_\varepsilon=1$ at $q_0$. For pure systems in two dimensions, the condition $X_\varepsilon=1$ is characteristic of free fermions, which have conformal dimension $\Delta_\eta=1/2$ ($\varepsilon\sim\eta\bar{\eta}$). It then follows from (\ref{phase}) that the zeros of $\rho_4$ should correspond to the values of $q$ for which $S=-1$ in the pure model. However, we know that, due to sign doubling, for such a value of $q$ there will also be a solution with $S=1$, so that it is sufficient to search for the values of $q$ for which the phase (\ref{S}) satisfies 
\EQ
\textrm{Im}\,S=(q-2)\textrm{Im}\,S_2=-(q-2)\rho\sin\varphi=0\,,\hspace{.6cm}\textrm{for} \ n=1\,,
\label{Im0}
\EN
This condition is satisfied for $q=2$, but also when $\rho=0$ or $\sin\varphi=0$. Inspection of the solutions listed in table~\ref{solutions} easily shows that the set of values of $q$ for which (\ref{Im0}) is satisfied precisely coincides with the set of zeros of $\rho_4$ exhibited by the solutions of section \ref{ct0}. 

The free fermionic realization in field theory is clear for $q=2$ (Ising) and, as an additional example, we illustrate it for $q=3$, which is a zero of $\rho_4$ for the solutions (\ref{eq:quenched_ue_sol1}) and (\ref{eq:quenched_ue_sol4}). For $q=3$ these two solutions (with the choice of lower signs) have $S=\cos\varphi=-1$, a result that in the pure model is only matched by solution I of table~\ref{solutions}. This solution can be identified with a line of fixed points with central charge 1, parametrized by $\varphi=2\pi\Delta_\eta$ \cite{DT}, and for $\Delta_\eta=1/2$ indeed corresponds to two free neutral fermions. 

A main difference between the scattering solutions for the pure case and those for the random case is that, while in the first case there are no fixed points for $q>(7+\sqrt{17})/2$, the second admits lines of fixed points extending to $q$ infinite. This result then appears as the explicit and exact manifestation of the findings of \cite{AW,HB} that in two dimensions quenched bond randomness can soften first order phase transitions into second order ones. More precisely, for the Potts model the rigorous result of \cite{AW} is that, if randomness leaves the couplings ferromagnetic, or just sufficiently so, the phase transition, which in the pure ferromagnet is first order for $q>4$, becomes second order up to $q$ infinite. We also know that the pure and random ferromagnetic critical lines meet at $q=2$, where randomness is marginally irrelevant \cite{DD}, so that the scattering solution for the random ferromagnetic line should have $\rho_4=0$ at $q=2$. Since the proof of \cite{AW} refers to integer values of $q$, these properties select two solutions\footnote{Solution (\ref{n0c}) has $\rho_4=0$ at $q=2$ but is not defined for $q=5$.} among those of section \ref{n0}, namely (\ref{n0b}) and (\ref{n0d}); the choice of lower signs in both cases ensures that $S_3$, the only physical amplitude for the pure model at $q=2$, takes the required value $-1$ at this point. On the other hand, numerical studies \cite{CJ} indicate that the ferromagnetic random critical line extends to real values of $q$, including the interval $q\in[3,4]$. If this property is required, solution (\ref{n0b}) remains as the only admissible. We notice that this solution is the $n\to 0$ limit of (\ref{eq:quenched_ue_sol2}), which for $n=1$ gives solution II of table~\ref{solutions}; (\ref{n0d}) instead, is the $n\to 0$ limit of (\ref{eq:quenched_ue_sol4}), which for $n=1$ gives solution III, namely the solution describing the ferromagnetic line of the pure model. 

\begin{figure}
\begin{center}
\includegraphics[width=10cm]{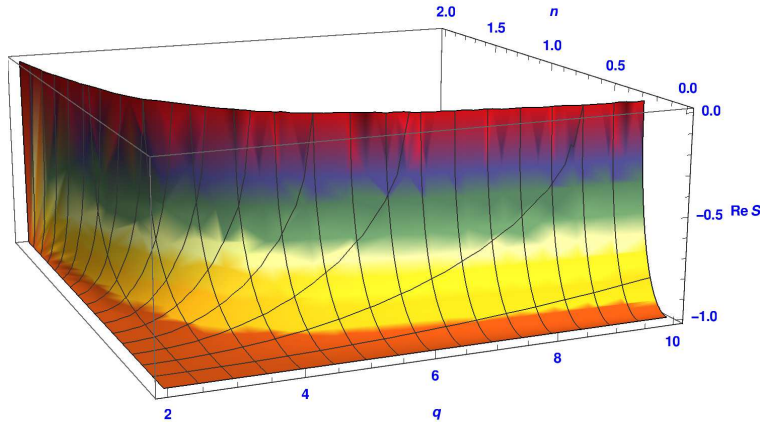}
\caption{Real part of the scattering phase (\ref{S}) for the solution (\ref{eq:quenched_ue_sol2}). It becomes $q$-independent in the limit $n\to 0$ corresponding to quenched disorder.
}
\label{S(q,n)}
\end{center} 
\end{figure}

A main property common to all solutions with $\cos\theta=0$ is that (\ref{S}) and (\ref{u3}) imply $\textrm{Im}\,S=0$ at $n=0$. This means that, while $S$ is $q$-dependent for $n$ generic, it becomes constant for $n=0$, i.e. precisely in the limit required for the random case. This is visible in figure~\ref{S(q,n)} for solution (\ref{eq:quenched_ue_sol2}), which has
\bea
\label{Smain}
S &=&\tfrac{1}{2}(q\rho_3 + i n(q-1)\rho_4\sin\theta)\\
&=&\exp[-i\sgn(n)\sgn(\sin\varphi)\Arccos(q\cos\varphi)]\,,\nonumber
\eea
where $\sgn(0)=1$.
This indicates that the symmetry sector of the state $\sum_{i,\gamma_i}A_{\alpha_i\gamma_i}A_{\gamma_i\alpha_i}$ becomes $q$-independent in the random case. This symmetry sector contains the fields which create the state $\sum_{i,\gamma_i}A_{\alpha_i\gamma_i}A_{\gamma_i\alpha_i}$, namely the $\mathbb{S}_q$-invariant fields like the energy density $\varepsilon$. This makes possible that the scaling dimensions of these fields remain constant along the random critical line. In particular, the correlation length critical exponent $\nu=1/(2-X_\varepsilon)$ can keep along the line the value 1 it takes at $q=2$. This value saturates the rigorous bound $\nu\geq 2/d$ for disordered systems in $d$ dimensions \cite{CCFS}. On the other hand, the spin field carries a representation of $\mathbb{S}_q$ symmetry and does not belong to the sector which becomes $q$-independent at $n=0$. As a consequence, the magnetization exponent $\beta$, which depends on the spin field scaling dimension, depends on $q$.

It is worth stressing how the $q$-independence of the scattering phase $S$ emerges, quite unexpectedly, as a property of the {\it exact} solution at $n=0$, and may not appear in an approximate calculation. This is why the scenario of a constant exponent $\nu$ is not in contradiction with the very weak $q$-dependence exhibited by leading order perturbative expansion in powers of $q-2$ for the random bond ferromagnet \cite{Ludwig,DPP}. The value $\nu=1$ is also consistent with the exact conjecture of \cite{AdAI} for $q\to\infty$. Concerning the numerical estimates of $\nu$ for the random bond ferromagnet, they appear to have the value $\nu=1$ within their error bars. The persistence of this circumstance at large values of $q$ \cite{CB2,OY,JP}, which are (also in principle\footnote{We recall that perturbation theory in powers of $q-2$ is not defined for $q>4$, where the transition in the pure model becomes first order.}) out of reach for the expansion in powers of $q-2$, calls for a non-perturbative mechanism, and the appearance of the $q$-independent amplitude $S$ in the exact scattering solution confirms the presence of such a mechanism.

It is also relevant to notice the following point concerning the multiscaling exponents associated to the moments of the energy-energy correlation function \cite{Ludwig,JL}. The numerical transfer matrix study of \cite{Jacobsen_multiscaling} obtained results consistent with a constant $\nu$, but exhibiting clear $q$-dependence for an energy multiscaling exponent. At first sight this circumstance can appear problematic, since both exponents fall in the $\mathbb{S}_q$-invariant sector of the theory. However, it must be noted that the state $\sum_{i,\gamma_i}A_{\alpha_i\gamma_i}A_{\gamma_i\alpha_i}$ associated to the scattering amplitude $S$ which becomes superuniversal at $n=0$ is both $\mathbb{S}_q$- and $\mathbb{S}_n$-invariant. On the other hand, it was shown in \cite{Ludwig} that, due to renormalization of products over different replicas, the fields entering the calculation of the generic energy multiscaling exponent are {\it not} $\mathbb{S}_n$-invariant\footnote{The issue does not arise in the calculation of the exponent associated to the first moment, corresponding to $\nu$, since it involves no product (see also \cite{Ludwig}).}. Hence, such exponents fall outside the symmetry sector associated to the amplitude $S$, and are expected to be $q$-dependent. It seems remarkable that the superuniversality of the scattering amplitude $S$ can give insight on the sensible difference in $q$-dependence that the data of \cite{Jacobsen_multiscaling} exhibit between $\nu$ and the multiscaling exponent.

\subsection{Solutions without weak disorder limit}
\label{strong_disorder}
Let us now pass to the two solutions with $\rho_6=0$. They are defined for all values of $q$ and are always strongly disordered ($\rho_4=1$). Solution (\ref{r60a}) is characterized by complete $q$-independence, a peculiarity which is not unexpected within the space of solutions. Indeed, for a disorder probability distribution
\EQ
P(J_{ij})=p\delta(J_{ij}-J)+(1-p)\delta(J_{ij})\,,\hspace{.5cm}J>0\,,
\label{dilute}
\EN
corresponding to a dilute ferromagnet, the energy receives contribution from clusters of spins connected by ferromagnetic bonds $J_{ij}=J$. At zero temperature, all the spins in such a cluster have the same color, different clusters are independently colored, and the total magnetization vanishes unless there is an infinite cluster; the latter appears when $p$ exceeds the random percolation threshold $p_c$. Hence, the dilute ferromagnet has a zero-temperature transition in the universality class of random percolation, no matter the value of $q$. It is then natural to associate this random percolation line to the $q$-independent solution (\ref{r60a}), for a suitable sign choice. 

It is relevant that this solution is defined, in particular, for $q=1$. Indeed, for $q\to 1$ the pure ferromagnet corresponds to random percolation with bond occupation probability $\tilde{p}=1-e^{-J/T}$ \cite{FK}. Hence, for $q\to 1$ the dilute ferromagnet corresponds to random percolation with bond occupation probability $p\tilde{p}$, and is critical for $p\tilde{p}=p_c$ ($\tilde{p}=1$ at $T=0$). Notice that none of the solutions admitting a weak disorder limit (those with $\cos\theta=0$) is defined for $q=1$. 

In the dilute ferromagnet, temperature can be shown to provide a relevant perturbation close to $T=0$ (see e.g. \cite{Cardy_book}). For $q>2$ this is consistent with a phase diagram with three fixed points along the paramagnetic/ferromagnetic phase boundary in the $p$-$T$ plane (left panel of figure~\ref{pd}): two unstable fixed points at the endpoints of the phase boundary (that of the pure model at $p=1$ and the percolative one at $T=0$) and an intermediate, stable fixed point corresponding to the solution with $q$-dependent disorder strength of the previous subsection. When $q\to 2$, the latter fixed point coalesces with the pure one and the renormalization group flow directly goes from the zero temperature to the pure fixed point. 

A qualitatively different scenario emerges for a disorder probability distribution (often called spin glass distribution)
\EQ
P(J_{ij})=p\delta(J_{ij}-J)+(1-p)\delta(J_{ij}+J)\,,\hspace{.5cm}J>0\,,
\label{glass}
\EN
allowing for ferromagnetic and antiferromagnetic bonds. A ferromagnetic/paramagnetic phase boundary is observed in the low-($1-p$) region of the $p$-$T$ plane (right panel of figure~\ref{pd}). A main difference with the dilute ferromagnet is the presence of an additional fixed point located on the phase boundary between a zero-temperature fixed point and the random (resp. pure) ferromagnetic fixed point for $q>2$ (resp. $q=2$). For $q=2$ a gauge symmetry \cite{Nishimori} allow to infer the presence of the additional fixed point, which is known as the Nishimori point. For $q>2$ Nishimori gauge symmetry does not hold but the Nishimori-like fixed point is still expected, and has been exhibited numerically for $q=3$ \cite{SGH}. 

\begin{figure}
\begin{center}
\includegraphics[width=9cm]{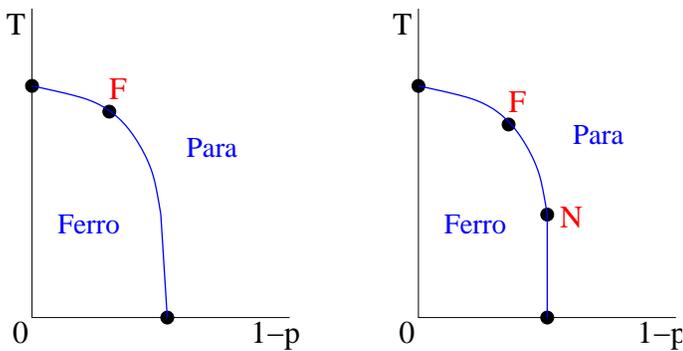}
\caption{Qualitative phase diagrams and expected fixed points (dots) in the $q>2$ two-dimensional Potts model with quenched bond distributions (\ref{dilute}) (left) and (\ref{glass}) (right); $N$ denotes the Nishimori-like fixed point. For $q=2$ the fixed point $F$ is absent and $N$ is the Nishimori point.
}
\label{pd}
\end{center} 
\end{figure}

Since the Nishimori-like fixed point is strongly disordered and cannot correspond to the scattering solutions with variable disorder strength (which have $\rho_4=0$ at $q=2$ whenever they are defined for this value of $q$), it should correspond to (\ref{r60a}) or (\ref{r60b}). While (\ref{r60a}) is completely $q$-independent, (\ref{r60b}) leads to $q$-independence of the scattering amplitude $\Sigma$ given by (\ref{Sigma}); once again, $q$-independence arises only at $n=0$, as can be seen from (\ref{r60n}). The symmetry sector associated to $\Sigma$ contains fields which create domain wall excitations. If one {\it assumes} that the duality relating the ferromagnetic and paramagnetic phases in the pure Potts model extends, at least in the continuum limit, to the replicated case, this symmetry sector contains the dual spin field, with the same scaling dimension $X_\sigma$ of the spin field; hence, $X_\sigma=\beta/\nu=\eta/2$ could be $q$-independent for the fixed point described by (\ref{r60b}), while $\nu$ is $q$-dependent. The numerically estimated exponents at the Nishimori(-like) fixed point are $\nu\approx 1.5$, $\eta\approx 0.18$ for $q=2$ \cite{MC,dQ,PHP,HPtPV,PtPV}, and $\nu=1.28-1.36$, $\eta=0.17-0.22$ for $q=3$ \cite{SGH}. These results seem to exclude solution (\ref{r60a}) (for which also $\nu$ is constant), but are compatible with (\ref{r60b}) and superuniversality of $\eta$. 

It is relevant to notice that a model possessing Nishimori gauge property (and a Nishimori critical point generalizing the Ising one) was introduced in \cite{NS} for the case of $\mathbb{Z}_q$ symmetry (cyclic permutations) and a specific realization of disorder. This model was studied numerically\footnote{See also \cite{OJ} for a study of the phase diagram.} for $q=3$ in \cite{JP2}, where the value $\eta=0.20-0.21$ was obtained at the Nishimori point. As observed in \cite{JP2}, it is an interesting question whether the $\mathbb{Z}_3$-invariant Nishimori point and the $\mathbb{S}_3$-invariant Nishimori-like point of \cite{SGH} belong to the same universality class. Should this be the case, the different values of $\eta$ at $q=2$ and $q=3$ would rule out superuniversality of this exponent. The simplest explanation would then be that the duality assumption on which superuniversality of $\eta$ relies does not hold, and the next task would be the identification of an exponent sensitive to the superuniversality of the amplitude $\Sigma$. 

Concerning the zero temperature fixed point for the disorder distribution (\ref{glass}), it is observed numerically that it is no longer in the percolation universality class \cite{PHP} and that it is stable under thermal perturbation \cite{SGH}, as required by the presence of the additional fixed point. Also the zero temperature fixed point is strongly disordered and should be described by (\ref{r60a}) or (\ref{r60b}), so that at least $\eta$ could be $q$-independent. The numerical results $\nu=1.35-1.45$, $\eta=0.18(1)$ for $q=2$ \cite{SGH,McMillan}, and $\nu=1.45-1.55$, $\eta=0.18(1)$ for $q=3$ \cite{SGH} are consistent with this scenario and point to $q$-dependent $\nu$, namely to (\ref{r60b}). However, the result $\eta\approx 0.13$ obtained in \cite{PHP,PtPV} at $q=2$ points instead to $q$-dependence also for $\eta$. In the latter case we would have solution (\ref{r60b}) with a violation of the duality assumption necessary for the superuniversality of $\eta$. 

We also recall that, even in the case in which the Nishimori-like and the zero temperature fixed points should turn out to correspond to the same scattering solution, this would not necessarily imply that they have equal critical exponents. This point was illustrated for the pure model in \cite{DT}, where it was shown that a single scattering solution accounts for the critical and tricritical ferromagnetic lines, as well as for the square lattice antiferromagnetic critical line.

\section{Conclusion}
In this paper we studied the exact scale invariant scattering solutions of two-dimensional quantum field theory with $\mathbb{S}_q\times\mathbb{S}_n$ symmetry. For $n$ (number of replicas) going to zero they describe the renormalization group fixed points of the $q$-state Potts model with quenched disorder. We found that all solutions exhibit, only in the limit $n\to 0$, $q$-independent sectors allowing for superuniversality (i.e. symmetry independence) of some critical exponents. We found two classes of solutions with non-vanishing disorder. The first class contains (up to an intrinsic sign doubling for all solutions) four solutions with disorder strength vanishing at points which are predictable using renormalization group arguments and the knowledge of the solutions for the pure case. All the solutions in this class are $q$-independent in the $\mathbb{S}_q$-invariant sector and allow for a superuniversal correlation length exponent $\nu$ and a $q$-dependent magnetic exponent $\eta=2\beta/\nu$. As already pointed out in \cite{random}, only one of these solutions has disorder strength vanishing at $q=2$ and is defined for {\it real} $q\geq 2$, as the line of fixed points followed numerically in \cite{CJ} in the random bond ferromagnet.   

The second class contains two solutions defined for all real values of $q$ and always exhibiting strong disorder. While one solution is completely $q$-independent and should account for the zero-temperature percolation fixed point in the bond dilute ferromagnet, the other is $q$-independent only in a $\mathbb{S}_q$ charged sector, a circumstance that, under a duality assumption, leaves room for superuniversality of the magnetic exponent $\eta$. The numerical results presently available at $q=2,3$ for the Nishimori(-like) fixed point in the model with $\pm J$ disorder do not yet exclude the latter scenario. The available data are not conclusive also for the zero-temperature fixed point in the same model, but seem to favor the solution with partial $q$-independence. When making these considerations, it must be taken into account that, as shown in \cite{DT} for the pure model, the same scattering solution can describe more than one critical line.

Concerning the determination of critical exponents from the scattering solutions, it was shown for pure models \cite{paraf,DT} how this is possible with sufficient insight about the underlying conformal field theory. This condition is not yet fulfilled for the case of quenched disorder. The very fact that random fixed points are conformally invariant has been difficult to substantiate in the past. The consistent picture provided by the results of \cite{random} and of the present paper now concretely supports conformal invariance, since the latter is responsible for complete elasticity of the scattering. The emergence of symmetry independent sectors, on the other hand, indicates that the conformal field theories describing random fixed points possess features which distinguish them from those known so far. 

We observed in section~\ref{variable_disorder} that the value $\nu=1$ for the random bond Potts ferromagnet saturates the rigorous result of \cite{CCFS} $\nu\geq 2/d$ for disorder systems. It is interesting to notice that precisely the condition $\nu\geq 1$ was argued in \cite{localization} to lead to interfacial wetting in near-critical two-dimensional pure systems at phase coexistence. In particular, for a Potts ferromagnet with a second order transition, slightly below critical temperature, boundary conditions can be chosen to generate an interfacial region separating a phase with predominant color 1 from a phase with predominant color 2. Interfacial wetting corresponds, for $q>2$, to an interfacial region being a double interface enclosing predominantly spins with color different from 1 and 2. If the argument of \cite{localization} extends to the random case, one should observe interfacial wetting at $q=3$ for the random ferromagnet, at variance with the pure case ($\nu=5/6$). While we are not aware of numerical data for $q=3$, a typical Monte Carlo configuration shown in \cite{FTM} for the $q=10$ random ferromagnet seems to suggest wetting.


\end{document}